\documentclass[aps,titlepage,preprint]{revtex4-2}
\usepackage[utf8]{inputenc}
\usepackage[margin=1in]{geometry}
\usepackage{amsmath,bm}
\usepackage{xcolor}
\usepackage{braket}
\usepackage{mathrsfs}
\usepackage{amsfonts}
\usepackage{epsfig,float}
\usepackage[normalem]{ulem}
\usepackage{amssymb,amsfonts}
\usepackage{accents}
\usepackage{graphicx}
\newcommand{\OmegaPBH}{\Omega_{\mathrm{PBH}}}

\begin{document}

\begin{titlepage}

\begin{center}

{\Large \bf  
Observable Signatures of No-Scale Supergravity in NANOGrav  
}

\vskip 0.5cm {\bf Spyros  Basilakos$^{a,b,c}$, Dimitri~V.~Nanopoulos 
$^{d,e,f}$,   Theodoros Papanikolaou$^{g,h,b}$, Emmanuel N. Saridakis$^{b,i,j}$,
Charalampos Tzerefos$^{k,b}$} 
\end{center}

{\small
\begin{quote}
\begin{center}
$^a$Academy of Athens, Research Center
for
Astronomy and Applied
Mathematics, Soranou Efesiou 4, 11527,
Athens,
Greece\\
$^b$National Observatory of Athens, Lofos
Nymfon, 11852 Athens, 
Greece\\
$^c$
School of Sciences, European University Cyprus, Diogenes 
Street, Engomi, 1516 Nicosia, Cyprus\\
$^d$ George P. and Cynthia W. Mitchell Institute for Fundamental
 Physics and Astronomy, Texas A\&M University, College Station, Texas 77843, 
USA\\
$^e$ Astroparticle Physics Group, Houston Advanced Research Center 
(HARC),
  Mitchell Campus, Woodlands, Texas 77381, USA \\
  $^f$ 
  Academy of Athens, Division of Natural Sciences,
Athens 10679, Greece \\
$^g$
Scuola Superiore Meridionale, Largo San Marcellino 10, 80138 Napoli, Italy
\\
 $^h$ 
Istituto Nazionale di Fisica Nucleare (INFN), Sezione di Napoli, Via Cinthia 21, 
80126 Napoli, Italy\\
 $^i$ CAS Key Laboratory for Researches in Galaxies and Cosmology, 
Department of Astronomy, University of Science and Technology of China, Hefei, 
Anhui 230026, P.R. China\\
 $^j$ 
 Departamento de Matem\'{a}ticas, Universidad Cat\'{o}lica del 
Norte, Avda.
Angamos 0610, Casilla 1280 Antofagasta, Chile\\
$^k$Physics Department,
University
of
Athens, Panepistemiopolis, Athens 157 83, Greece
\end{center}%

\end{quote}}

%
 
\vspace{0.2cm}
\centerline{\bf Abstract}
\bigskip 
In light of NANOGrav data we provide for the first time possible
observational signatures of Superstring theory.
Firstly, we work with inflection-point inflationary potentials 
naturally realised
within Wess-Zumino type no-scale Supergravity, which give rise
to the formation of microscopic primordial black holes
(PBHs) triggering an early matter-dominated era (eMD) and
evaporating before Big Bang Nucleosythesis
(BBN). Remarkably, we obtain an abundant production
of primordial gravitational waves (PGW)
at the frequency ranges of nHz, Hz and kHz
and in strong agreement with Pulsar Time Array (PTA) GW data. 
This PGW background could serve as a compelling observational
signature for the presence of quantum gravity via no-scale Supergravity.

%
\vspace{0.3cm}%
%


%
%
\end{titlepage}


\pagestyle{plain} \baselineskip0.75cm 

 After 50 years of Superstring theory one of the most fundamental problems is the following: {\it Is it ever possible to macrospopically measure its effects?} In this essay we argue that the  current and future  gravitational-wave (GW) data provide the necessary platform in order  to answer this question. Specifically, recent findings  from the 15-year pulsar timing array data release by the 
NANOGrav Collaboration suggest compelling evidence supporting the existence of a 
low-frequency gravitational-wave (GW) background \cite{NANOGrav:2023gor}, being explained within a plethora of cosmological setups ~\cite{Bringmann:2023opz,Choudhury:2023kam,Wu:2023pbt,Balaji:2023ehk,Gouttenoire:2023nzr,Bhaumik:2023wmw,Franciolini:2023wjm,Franciolini:2023pbf,Inomata:2023drn,Datta:2023vbs,Choudhury:2023fwk,Huang:2023chx,Choudhury:2024dzw}. In this essay, we propose a robust mechanism for the generation of such a signal 
within the framework of no-scale Supergravity 
\cite{Cremmer:1983bf,Ellis:1983sf,Ellis:1984bm,Lahanas:1986uc,Freedman:2012zz}, 
which represents the low-energy limit of Superstring theory 
\cite{Witten:1985xb,Dine:1985rz,Antoniadis:1987zk,Horava:1996vs}. This framework naturally 
yields a Starobinsky-like inflation \cite{Ellis:2013nxa,Kounnas:2014gda} that 
exhibits the desired observational characteristics, alongside a coherent 
particle and cosmological phenomenology elucidated within the 
superstring-derived flipped SU(5) no-scale Supergravity 
\cite{Antoniadis:2020txn,Antoniadis:2021rfm}.  
 
 The present study is novel since it opens a new avenue in understanding the nature of the fundamental
 theory of gravity.  In our current investigation, we use the Wess-Zumino no-scale 
Supergravity \cite{Ellis:2013xoa}, and within  this context, we explore 
naturally occurring inflection-point single-field inflationary potentials. 
These potentials hold the capacity to engender the formation of microscopic 
Primordial Black Holes (PBHs). These PBHs exhibit masses below 
$10^9\mathrm{g}$, potentially instigating early matter-dominated eras (eMD) 
preceding Big Bang Nucleosynthesis (BBN)~\cite{Garcia-Bellido:1996mdl, Hidalgo:2011fj, Suyama:2014vga, Zagorac:2019ekv} and being associated with a very rich phenomenology~\cite{ Barrow:1990he, Bhaumik:2022pil, Banerjee:2022xft, Papanikolaou:2023oxq}. 
Furthermore, we investigate the stochastic gravitational-wave (GW) signals 
induced by second-order gravitational interactions stemming from inflationary 
adiabatic perturbations. Additionally, we explore GW signals arising from 
isocurvature-induced adiabatic perturbations due to Poisson fluctuations in the 
number density of PBHs. For interesting reviews on PBHs see here~\cite{Carr:2009jm,Escriva:2022duf,Choudhury:2024aji}

Remarkably, our analysis reveals a 
three-peaked induced GW signal spanning the 
frequency ranges of $\mathrm{nHz}$, $\mathrm{Hz}$, and $\mathrm{kHz}$, aligning 
closely with the recently published Pulsar Timing Array (PTA) GW data \cite{NANOGrav:2023gor}. The 
simultaneous detection of these three GW peaks by current and future GW 
detectors could serve as a compelling observational signature, a smoking gun for the presence 
of no-scale Supergravity.

\textit{No-scale Supergravity} -- In the most general $N=1$ supergravity theory, three essential functions come 
into play, namely the  K\"ahler potential $K$ (which is a hermitian function of the scalar field), the superpotential $W$ and the 
function $f_{ab}$, which are holomorphic functions of the fields.  
We examine a no-scale supergravity model featuring two chiral superfields, $T$ 
and $\varphi$, which parameterize the noncompact $SU(2,1)/SU(2) \times U(1)$ 
coset space. The  K\"ahler potential for this model is given by 
\cite{Ellis:1984bm,Nanopoulos:2020nnh}:
\begin{equation}
K= -3 \ln \left[ T +\bar{T} -\frac{\varphi \bar{\varphi}}{3}+a  e^{-b(\varphi+ 
\bar{\varphi})^2}(\varphi+\bar{\varphi} )^4 \right],
\label{a1}
\end{equation}
where $a$ and $b$ are real constants.
The simplest globally supersymmetric model, known as the Wess-Zumino model, 
involves a single chiral superfield $\varphi$. It includes a mass term 
$\hat{\mu}$ and a trilinear coupling $\lambda$, with the corresponding 
superpotential given by \cite{Ellis:2013xoa}:
\begin{equation}
W= \frac{\hat \mu}{2} \varphi^2 - \frac{\lambda}{3} \varphi^3.
\label{1a}
\end{equation}
In the limit $a=0$, and by identifying the $T$ field with the modulus field and 
the $\varphi$ field with the inflaton field, one can derive a class of no-scale 
theories that yield Starobinsky-like effective potentials 
\cite{Ellis:2013xoa,Ellis:2013nxa}. The potential is evaluated along the real 
inflationary direction aka for $T= \bar{T}= \frac{c}{2}$ and  $\mathrm{Im}\varphi=0 $ with $c$ a constant) ,  
with $\lambda / \mu = 1/3$ and $\hat{\mu}=\mu \sqrt{c/3}$.  Note that by defining $ \phi \equiv \mathrm{Re}\varphi $ and transforming it to a new field $\chi$ through
$
\varphi= \sqrt{3 c}  \tanh \left( \frac{\chi }{\sqrt{3}} \right)$, one
recovers the Starobinsky potential, namely $ V (\chi)=\frac{\mu^2}{4} \left(1- 
e^{-\sqrt{\frac{2}{3}} , { \chi} } \right)^2 $.

 \textit{Inflationary Dynamics} -- 
 Now, let's examine the inflationary dynamics both at the background and 
perturbative levels. Working within a flat Friedmann-Lemaître-Robertson-Walker 
(FLRW) background, the metric is represented by $\mathrm{d}s^2 = -\mathrm{d}t^2 
+a^2(t)\mathrm{d}x^i\mathrm{d}x_i$, with $a(t)$ the scale factor, and the Friedmann equations take the 
familiar form:
\begin{eqnarray}
&&H^2 = \frac{1}{3 }\left[\frac{\dot{\chi}^2}{2} + 
V\left(\phi(\chi)\right)\right], \\
&&\dot{H} = - \frac{\dot{\chi}^2}{2}.
\end{eqnarray}
 Due to the presence of an inflection point 
in the inflationary potential, we obtain  a transient 
ultra-slow-roll (USR) period. During this USR phase, the non-constant mode of the
curvature fluctuations, which would typically decay in standard slow-roll 
inflation, instead grows exponentially, thereby enhancing the curvature power 
spectrum at small scales and leading to the formation of Primordial Black Holes 
(PBHs). This phenomenon is a direct consequence of the extended  K\"ahler
potential. Moreover, for a suitable choice of the theoretical parameters (see below)  
the aforementioned inflationary potential   
yields 
a spectral index $n_\mathrm{s}\simeq 0.96$ and a tensor-to-scalar ratio 
$r<0.04$, in excellent agreement with Planck data \cite{Planck:2018vyg}.
  
\begin{figure}[t!]
\begin{centering}
\includegraphics[width=0.485\textwidth]{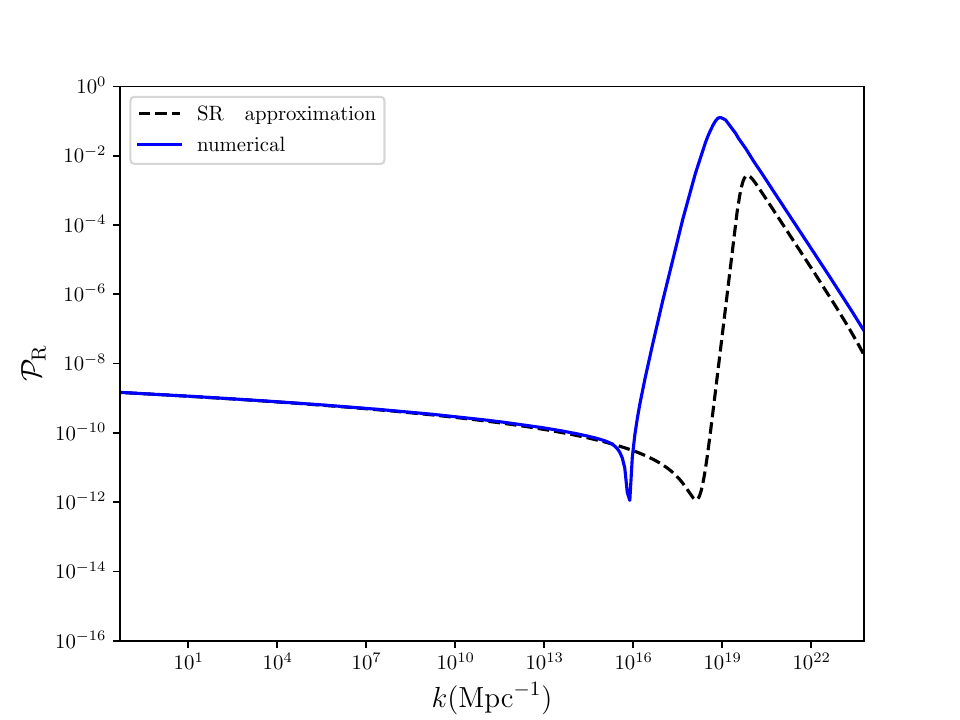}
\includegraphics[width=0.485\textwidth, clip=true]{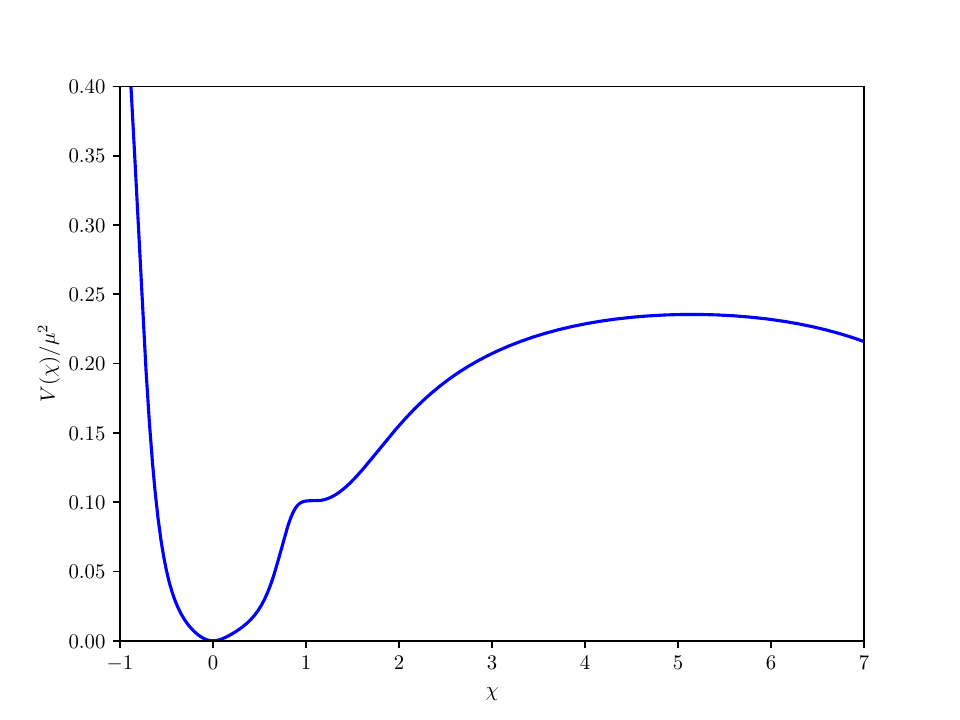}
\caption{{\it{Left Panel: The curvature power spectrum
$\mathcal{P}_\mathcal{R}(k)$  as a function of the wavenumber $k$, for 
$a=-1$, $b=22.35$, 
$c=0.065$, $\mu = 2\times 10^{-5}$, $\lambda/\mu = 0.3333449$ and $\phi_0 = 
0.4295$ in Planck 
units. The black 
dashed curve represents the slow-roll (SR) approximation for 
$\mathcal{P}_\mathcal{R}(k)$, 
while the blue solid curve is the exact one after the numerical integration 
of the Mukhanov-Sasaki equation. Right Panel: The potential $V(\chi)$ in terms of $\chi$ in Planck units for the same choice of fiducial parameters.}}}
\label{fig:P_RR}
\end{centering}
\end{figure}

Now, let us concentrate on the perturbative level, focusing on the comoving 
curvature 
perturbation defined as:
$
\mathcal{R} \equiv \Phi + \frac{H}{\dot{\chi}}\delta \chi,
$
where $\Phi$ represents the Bardeen potential, or otherwise the first order scalar metric perturbation. We 
extract the Mukhanov-Sasaki (MS) equation, and we focus on the curvature power 
spectrum defined as
\begin{equation}
\mathcal{P}_{\mathcal{R}}(k) \equiv 
\left(\frac{k^{3}}{2\pi^{2}}\right)|\mathcal{R}{_k}|^{2}.
\label{eq:P_R}
\end{equation}
We numerically then integrate the MS equation under the Bunch-Davies 
vacuum initial conditions on subhorizon scales and in  Fig. \ref{fig:P_RR}, 
we present  $\mathcal{P}_\mathcal{R}(k)$ as well as $V(\chi)$
for some fiducial values of the theoretical parameters involved, namely $a=-1$, 
$b=22.35$, $c=0.065$, $\mu = 2\times 10^{-5}$, and $\lambda/\mu = 0.3333449$. 
(Note that the value $\lambda/\mu = 1/3$ alongside $a=0$ corresponds to the 
Starobinsky model). The initial value of the $\phi$ field was taken as $\phi_0 = 
0.4295$ in Planck units.

Remarkably, as shown in Fig. \ref{fig:P_RR},   $\mathcal{P}_\mathcal{R}(k)$ peaks 
at $k\mathrm{peak}\sim 10^{19}\mathrm{Mpc}^{-1}$, corresponding to a PBH mass 
forming in the radiation-dominated era (RD) of the order of 
$M_\mathrm{PBH}=17M_\odot\left(\frac{k}{10^6\mathrm{Mpc}^{-1}}\right)^{-2}\sim 
10^8\mathrm{g}$ \cite{Carr:2020gox}, which evaporates at around $1\mathrm{MeV}$, 
i.e., BBN time.

  \textit{The primordial black hole gas} -- 
 PBHs originate from the collapse of local overdensity regions when the energy 
density contrast of the collapsing overdensity surpasses a critical threshold 
$\delta_\mathrm{c}$ \cite{Harada:2013epa, Musco:2020jjb}. In our analysis
  we have 
accounted for the non-linear relationship between the energy density contrast 
$\delta$ and the comoving curvature perturbation $\mathcal{R}$, leading to an 
inherent primordial non-Gaussianity of the $\delta$ field 
\cite{DeLuca:2019qsy,Young:2019yug} \footnote{We need to note here that the presence of an inflection point in the inflationary potential leads unavoidably to a deviation from the slow-roll attractor behaviour, which is connected with potentially significant non-Gaussianities~\cite{Martin:2012pe}, hence affecting  the statistics of the curvature perturbation $\mathcal{R}$. In particular, a highly non-Gaussian tail of the probability distribution function (PDF) of $\mathcal{R}$ can alter (enhance/suppress) significantly the PBH abundances~\cite{Atal:2018neu,Matsubara:2022nbr,Gow:2022jfb,Ferrante:2022mui} and potentially, depending on the model, modify the spectral shape of the induced GW signal~\cite{Cai:2018dig,Papanikolaou:2024kjb}. However, the amplitude of induced GWs being dependent mainly on the variance of the PDF of $\mathcal{R}$ will change mildly~\cite{Nakama:2016gzw,Garcia-Bellido:2017aan,Cai:2018dig,Unal:2018yaa,Garcia-Saenz:2022tzu,Abe:2022xur}. In this work,  we restrict ourselves only on the effect of non-Gaussianity in the density contrast which originates from the non-linear relation between the curvature perturbation $\mathcal{R}$ and the density contrast $\delta$.}. 
 
\begin{figure*}[t!]
\begin{centering}
\includegraphics[width=0.495\textwidth]{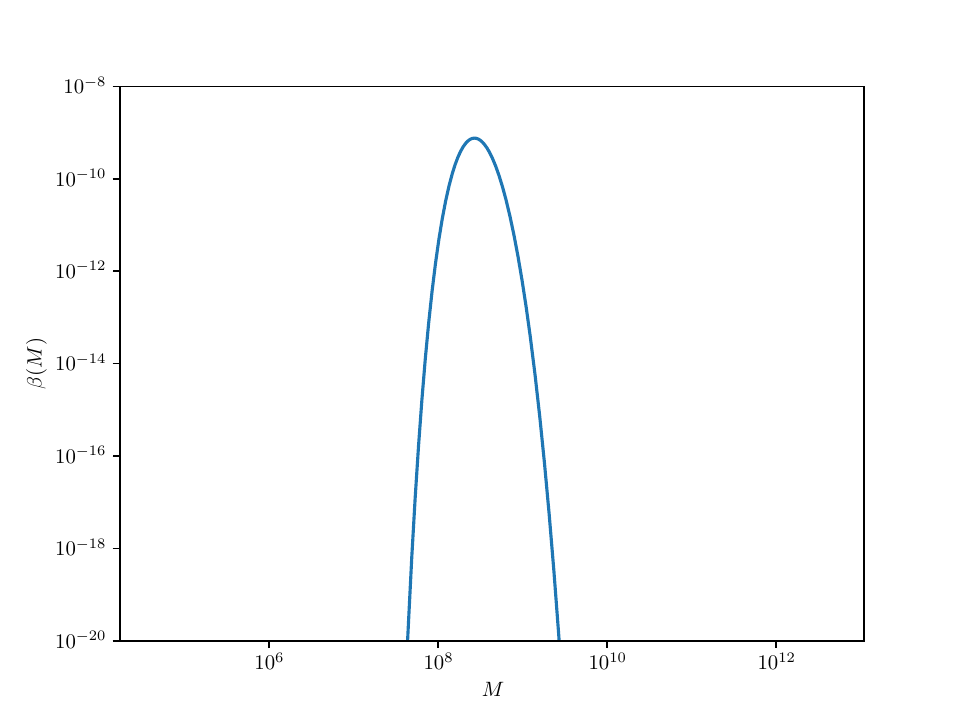}
  \includegraphics[width=0.495\textwidth, clip=true]
                  {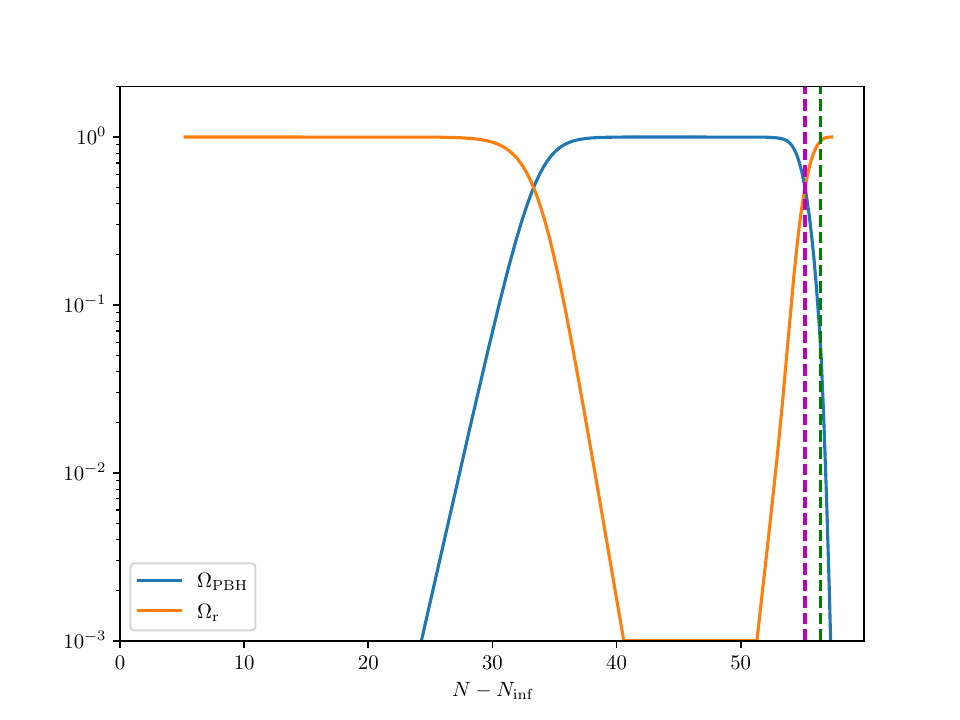}
  \caption{{\it{Left Panel: The PBH mass function $\beta(M)$ 
\cite{Papanikolaou:2022chm} at formation. Right Panel: The dynamical evolution 
of the 
background PBH and radiation
energy densities as a function of the e-folds passed from the end of inflation. 
The magenta 
vertical dashed line denotes 
the time of the onset of the radiation-dominated era, namely when 
$\Omega_\mathrm{r} = 0.5$, whereas the green dashed vertical line stands for 
the time when $\Omega_\mathrm{r} = 0.95$, namely when we are fully back to the 
radiation-dominated Universe. For both panels, we used $a=-1$, $b=22.35$, 
$c=0.065$, $\mu = 2\times 10^{-5}$, $\lambda/\mu = 0.3333449$ and $\phi_0 = 
0.4295$ in Planck 
units. 
}}}
\label{fig:Omega_PBH}
\end{centering}
\end{figure*}

  Operating within the framework of Wess-Zumino type no-scale supergravity 
  we observe an amplified curvature power spectrum, which, 
while broader compared to the Dirac-monochromatic case (see  Fig. 
\ref{fig:P_RR}), still exhibits sharp features, naturally giving rise to nearly 
monochromatic distributions of PBH masses  (see the left panel of 
Fig. \ref{fig:Omega_PBH}). Consequently, we anticipate a ``population'' of PBHs 
with varying masses lying within the range $[10\mathrm{g},10^9\mathrm{g}]$, 
thus undergoing evaporation prior to BBN \cite{Kawasaki:1999na}. Nonetheless, 
the majority of these PBHs are likely to possess a common mass corresponding to 
the peak of the primordial curvature power spectrum.

After performing a redshift evolution and accounting for the effect of Hawking evaporation we 
obtain the energy densities of PBHs and the radiation background, as 
illustrated in the right panel of Fig. \ref{fig:Omega_PBH}. As observed, the 
abundance of PBHs increases with time due to cosmic expansion. At early times 
when $\OmegaPBH\ll 1$ and Hawking radiation is negligible, 
$\Omega_\mathrm{PBH}$ behaves as $\propto a$, dominating the Universe's energy 
budget for a brief period. However, at some point, Hawking evaporation becomes 
the dominant factor in the dynamics of $\OmegaPBH$, causing the PBH abundance to 
decrease.

  \textit{Scalar induced gravitational waves} -- 
Let's proceed with the derivation of gravitational waves induced by second-order 
gravitational interactions stemming from first-order curvature perturbations.  
We begin our analysis in the Newtonian gauge, a commonly adopted framework in 
studies related to secondary-induced gravitational waves (SIGWs).      
  After a lengthy yet straightforward calculation, one obtains a tensor power 
spectrum $\mathcal{P}_{h}(\eta,k)$. 
Consequently, we can extract the gravitational wave (GW) spectral abundance, 
defined as the GW energy density per logarithmic comoving scale. 
Finally, considering the radiation energy density $\rho_r = 
\frac{\pi^2}{30}g_{*\mathrm{\rho}}T_\mathrm{r}^4$ and the temperature of the 
primordial plasma $T_\mathrm{r}$ scaling as $T_\mathrm{r}\propto 
g^{-1/3}_{*\mathrm{S}}a^{-1}$, the GW spectral abundance at our present epoch is 
given by:
\begin{equation}
\Omega_\mathrm{GW}(\eta_0,k) = 
\Omega^{(0)}_r\frac{g_{*\mathrm{\rho},\mathrm{*}}}{g_{*\mathrm{\rho},0}} 
\left(\frac{g_{*\mathrm{S},\mathrm{0}}}{g_{*\mathrm{S},\mathrm{*}}}\right)^{4/3} 
\Omega_\mathrm{GW}(\eta_\mathrm{*},k),
\label{Omega_GW_RD_0}
\end{equation}
where $g_{*\mathrm{\rho}}$ and $g_{*\mathrm{S}}$ represent the energy and 
entropy relativistic degrees of freedom. It's worth noting that the reference 
conformal time $\eta_\mathrm{*}$, in the case of an instantaneous transition 
from the eMD to the lRD era, should be of $\mathcal{O}(1)\eta_r$ 
\cite{Inomata:2019ivs,Inomata:2020lmk}. In the scenario of a gradual transition, 
$\eta_\mathrm{*}\sim (1-4)\eta_\mathrm{r}$ is necessary for $\Phi$ to have 
sufficiently decayed and for the tensor modes to be treated as freely 
propagating gravitational waves \cite{Inomata:2019zqy,Papanikolaou:2022chm}.

\begin{figure}[h!]
\begin{center}
\includegraphics[width=0.52\textwidth]{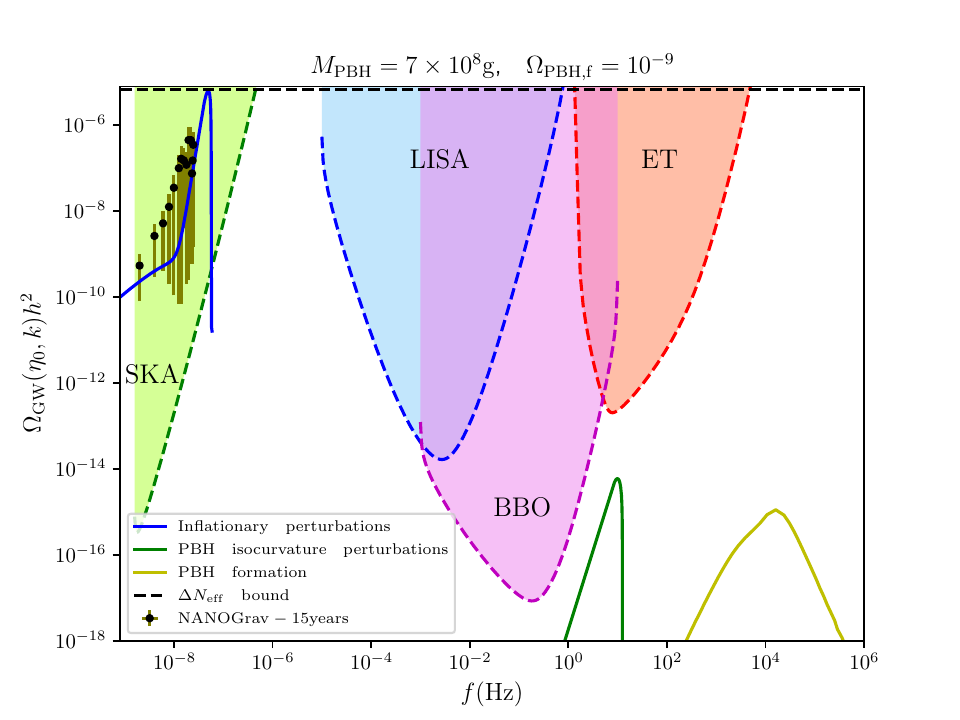}
\caption{ 
{\it{The stochastic three-peaked GW background induced by inflationary adiabatic 
(solid blue and yellow curves) and PBH isocurvature perturbations (solid green 
curve) arising from Wess-Zumino no-scale Supergravity with the extended K\"ahler 
potential (\ref{a1}) for $a=-1$, $b=22.35$, $c=0.065$, $\mu = 2\times 10^{-5}$, 
$\lambda/\mu = 0.3333449$ and $\phi_0 = 0.4295$ in Planck units and 
corresponding to $M_\mathrm{PBH}=7\times 10^8\mathrm{g}$ and 
$\Omega_\mathrm{PBH,f}= 10^{-9}$. On the top of our theoretical prediction for 
the induced stochastic GW background we show the  15-year NANOGrav GW data, as 
well as the sensitivities of SKA~\cite{Janssen:2014dka}, 
LISA~\cite{LISACosmologyWorkingGroup:2022jok,Karnesis:2022vdp}, BBO~\cite{Harry:2006fi} and 
ET~\cite{Maggiore:2019uih} GW experiments. In the horizontal black dashed line, 
we show also the upper bound on $\Omega_\mathrm{GW,0}h^2\leq 6.9\times 10^{-6}$ 
coming from the upper bound constraint on $\Delta N_\mathrm{eff}$ from CMB and 
BBN observations~\cite{Smith:2006nka}.} 
}}
\label{fig:GW_signals}
\end{center}
\end{figure}

\begin{figure}[h!]
\hspace{-0.8cm}
\includegraphics[width=0.52\textwidth]{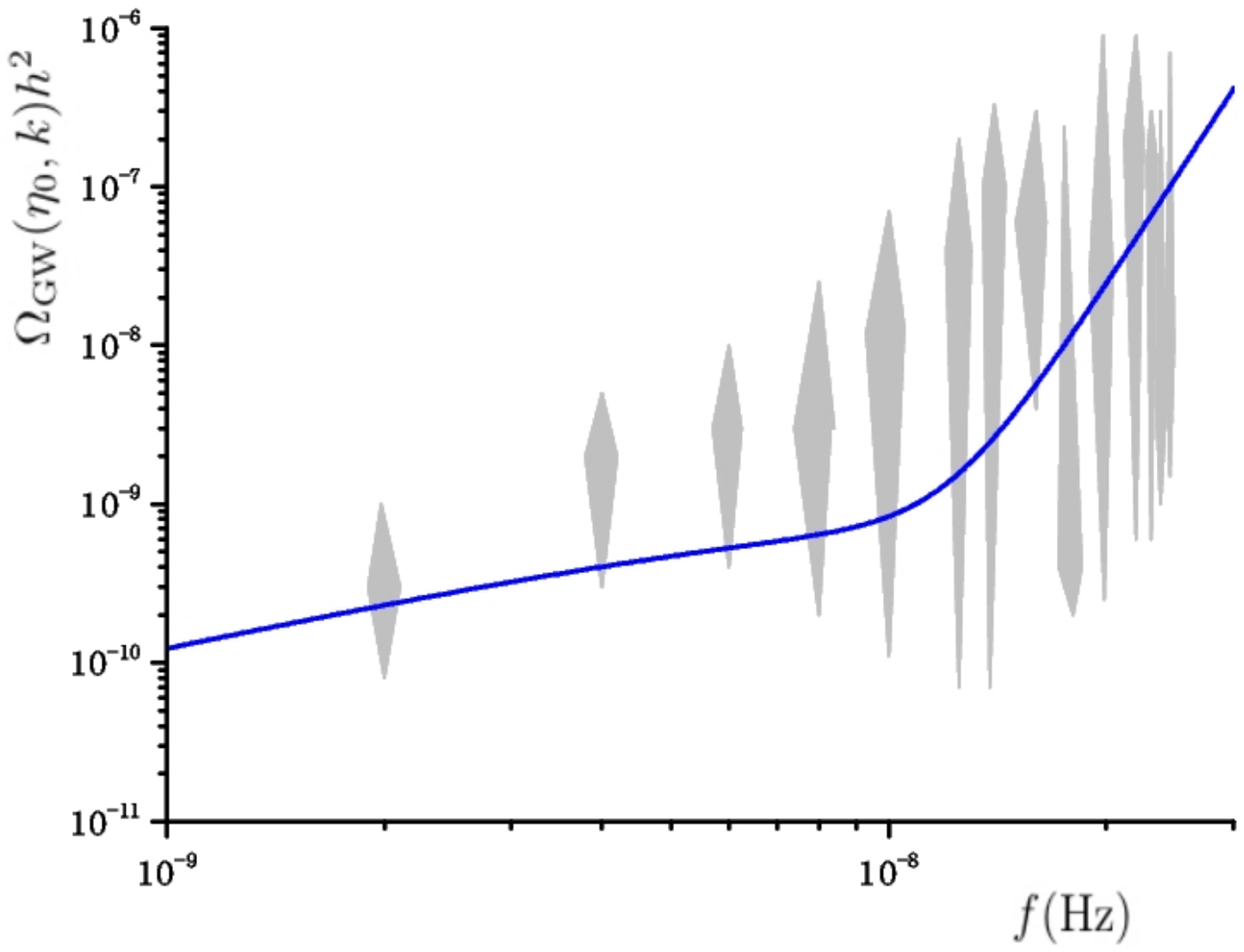}
\caption{ {\it{The stochastic GW background (GW density 
parameter as a function of the frequency) induced by resonantly amplified 
inflationary perturbations and 
arising from Wess-Zumino no-scale Supergravity with the extended K\"ahler 
potential (\ref{a1}) for  $a=-1$, $b=22.35$, 
$c=0.065$, $\mu = 2\times 10^{-5}$, $\lambda/\mu = 0.3333449$ and $\phi_0 = 
0.4295$ in Planck units - corresponding to $M_\mathrm{PBH}=7\times 
10^8\mathrm{g}$ and $\Omega_\mathrm{PBH,f}= 10^{-9}$ - on top of the 15-year 
NANOGrav data.
}}}
\label{fig:NANOGrav_comparison_figure}
\end{figure}

\textit {The relevant gravitational-wave sources} -- 
We can now focus on the different sources of gravitational waves (GWs) 
considered within this work.  The first GW source is related to GWs induced by the 
enhanced primordial curvature power spectrum around the PBH scale, namely around 
$k=10^{19}\mathrm{Mpc}^{-1}$, which is associated with PBH formation.   In the 
end, for our fiducial choice of the inflationary parameters involved, the GW 
signal associated with PBH formation peaks in the kHz frequency range [See the 
yellow solid curve in Fig. \ref{fig:GW_signals}]. 
Regarding the second GW peak at nHz, it is related to the resonant amplification 
of the curvature perturbation on scales entering the cosmological horizon 
during the eMD
[see~\cite{Inomata:2019ivs,Domenech:2021ztg} for more details.]  
  Remarkably, this second peak that corresponds to scales much larger than the 
PBH scale peaks at the $\mathrm{nHz}$ frequency range and is in strong 
agreement with the NANOGrav/PTA data \cite{NANOGrav:2023gor} - see the blue solid curve in Fig. 
\ref{fig:GW_signals} as well as Fig. \ref{fig:NANOGrav_comparison_figure} where 
we have zoomed in the NANOGrav frequency range.

Finally, we consider the GW spectrum induced by the gravitational potential of 
our primordial black hole (PBH) population itself. Assuming that PBHs are 
randomly distributed at formation time (i.e., they have Poisson 
statistics)~\cite{Desjacques:2018wuu,MoradinezhadDizgah:2019wjf}, their energy 
density is inhomogeneous while the total background radiation energy density is 
homogeneous. Therefore, the PBH energy density perturbation can be described by 
an isocurvature Poisson fluctuation~\cite{Papanikolaou:2020qtd}, which in the 
subsequent PBH domination era will be converted into an adiabatic curvature 
perturbation associated with a PBH gravitational potential $\Phi_\mathrm{PBH}$.  
  As one can see from the green solid line of Fig. \ref{fig:GW_signals}, for 
our fiducial choice of the inflationary parameters at hand, this GW signal lies 
within the $\mathrm{Hz}$ frequency range with an amplitude of the order of 
$10^{-14}$, being close to the sensitivity bands of the Einstein Telescope 
(ET)~\cite{Maggiore:2019uih} and Big Bang Observer (BBO)~\cite{Harry:2006fi}.

At this point we should mention that our choice of parameters is not unique, therefore a full parameter space analysis is required to exhaustively investigate the observational signatures of our model. However, given the very delegate dependence of the PBH formation on our parameters, such an analysis is beyond the scope of this work and will be performed elsewhere.

Lastly, it is noteworthy that since gravitational waves (GWs) generated before 
BBN can act as an extra relativistic component, they will 
contribute to the effective number of extra neutrino species $\Delta 
N_\mathrm{eff}$, which is severely constrained by BBN and Cosmic Microwave 
Background (CMB) observations as $\Delta N_\mathrm{eff} < 
0.3$~\cite{Planck:2018vyg}. This upper bound constraint on $\Delta 
N_\mathrm{eff}$ is translated to an upper bound on the GW amplitude, which 
reads as~\cite{Smith:2006nka,Caprini:2018mtu} $\Omega_\mathrm{GW,0}h^2 \leq 6.9 
\times 10^{-6}$. This upper bound on 
$\Omega_\mathrm{GW}$ is shown with the horizontal black dashed line in Fig. 
\ref{fig:GW_signals}.

{\textit Conclusions} -- 
In this essay, we showed that no-scale Supergravity, being the low-energy 
limit of Superstring theory, seems to accomplish three main achievements. 
Firstly, it provides a successful Starobinsky-like inflation realization with 
all the desired observational predictions regarding $n_\mathrm{s}$ and 
$r$~\cite{Antoniadis:2020txn,Ellis:2013xoa}. Secondly, it can naturally lead to 
inflection-point inflationary potentials giving rise to sharp mass 
distributions of microscopic PBHs triggering an eMD era before BBN, and thirdly 
it can induce through second order gravitational interactions a distinctive 
three-peaked GW signal.

In particular, working within the context of Wess-Zumino no-scale Supergravity 
we found i) a $\mathrm{nHz}$ GW signal induced by enhanced inflationary 
adiabatic perturbations and resonantly amplified due to the sudden transition 
from a eMD era driven by ``no-scale" microscopic PBHs to the standard RD era, 
being as well within the error-bars of the recently PTA GW data, ii) a 
$\mathrm{Hz}$ GW signal induced by the PBH isocurvature energy density 
perturbations and close to the GW sensitivity bands of ET and BBO GW experiments 
and iii) a $\mathrm{kHz}$ GW signal associated to PBH formation. Remarkably, 
a simultaneous detection of all three $\mathrm{nHz}$, $\mathrm{Hz}$ and 
$\mathrm{kHz}$ GW peaks can constitute a potential observational signature for 
no-scale Supergravity.
 
It should be stressed that No-Scale Supergravity, the low-energy limit of Superstring theory \cite{Witten:1985xb,Dine:1985rz,Antoniadis:1987zk,Horava:1996vs} is an essential element in constructing Superstring models incorporating particle physics and cosmology \cite{Ellis:2013nxa,Kounnas:2014gda}. As such, any strengthening of the consequences of No-Scale Supergravity
in the light of PTA GW data, discussed here, will reflect very favourably as the first time ever possible observable signature of  Superstring theory.


\vspace{1cm}

\bibliography{ref}

\end{document}